\documentclass{ws-p8-50x6-00}
\usepackage{amsfonts}
\usepackage{amsbsy}
\input{definitionen.tex}

\begin{document}
\title{A rigorous path-integral formula for quantum\\spin dynamics via
  planar Brownian motion 
}

\author{Bernhard Bodmann}

\address{Department of Mathematics, University of Florida, 358 Little Hall,
Gainesville,\\ Fl 32611, USA}

\author{Hajo Leschke \lowercase{and} Simone Warzel}

\address{Institut f\"ur Theoretische Physik, Universit\"at Erlangen-N\"urnberg,
Staudtstr. 7, D-91058 Erlangen, Germany}  


\maketitle

\abstracts{
Adapting ideas of Daubechies and Klauder
we derive a continuum path-integral formula for the
time evolution generated by a spin Hamiltonian.
For this purpose we identify the finite-dimensional spin Hilbert 
space with the ground-state eigenspace of a suitable Sch\"odinger operator 
on $\mbox{\rm L}^2({\mathbb{R}}^2)$, the Hilbert space of 
square-integrable functions on the Euclidean plane ${\mathbb{R}}^2$,
and employ the Feynman-Kac-It\^o
formula.}

\section{Introduction}
Even 50 years after the appearance of Feynman's celebrated $\mbox{paper}^1$
which introduced 
the path-integral formalism into quantum theory in a heuristic but convincing 
manner, there is no general consensus on how to treat a quantum spin 
within this framework. To our knowledge, among the various approaches the only rigorous
expression for the dynamics of a quantum spin in terms of an integral over continuous
paths is due to Daubechies and Klauder.\cite{DaKl85}
They were able to write the coherent representation of the unitary time-evolution 
operator of a spin with definite quantum number
as the ultra-diffusive limit of a 
well-defined integral over Brownian-motion paths on the two-sphere.

The present contribution summarizes a recent $\mbox{work}^3$, where we have shown that
one may equally well employ Brownian-motion paths on the 
Euclidean plane. In this way 
a closer contact to symbolic continuum path-integral formulas widely discussed in the 
recent $\mbox{literature}^{4-7}$ is established. 
One may hope that the wealth of analytical tools
associated with planar Brownian motion helps clarifying some subtle points there.
The presented approach might also be of interest in the context of quantizing 
constrained systems.  
\section{Basic definitions, result, and comments}
%
%
%
We consider a single {\bf spin} with fixed {\bf quantum number} 
$j \in \left\{0, 1/2, 1, 3/2, \dots \right\}$. 
It is given by three operators ${\cal J_+}, {\cal J_-}$, and $ {{\cal J}_3}$ which are
viewed as acting on the $(2j+1)$-dimensional complex Hilbert 
space ${\mathbb{C}}^{2j+1}$ and   
obey the usual commutation relations
$
   [{\cal J_+}, {\cal J_-}] = 2 {{\cal J}_3} $, 
$ [{\cal J}_3 , {\cal J_\pm}] = \pm {\cal J_\pm} 
$.
%
%
Non-normalized {\bf coherent vectors}
        \begin{equation}
                \ket{z} := [(2j+1)/\pi]^{1/2}
                        \left(1 + |z|^2 \right)^{-j-1} e^{z {\cal J_+}} \ket{j,-j}
\, , \quad  z \in {\mathbb{C}}\, ,
        \end{equation}
in ${\mathbb{C}}^{2j+1}$ 
are parametrized by complex numbers. 
Henceforth,  $z^*$ will refer to
their complex conjugates, $z_1:=(z+z^*)/2$ and $z_2:= (z-z^*)/2i$
to their real and imaginary parts. 
A  normalized {\bf spin-down vector} $\ket{j,-j} \in {\mathbb{C}}^{2j+1}$, 
obeying
$
        {\cal J_-} \ket{j,-j} = 0
$ and
$
        \braket{j,-j}{j,-j} = 1 \, ,
$
serves as the reference vector.

Every {\bf spin Hamiltonian} ${\mathcal H}$, 
a (self-adjoint) operator on ${\mathbb{C}}^{2j+1}$, is polynomial
in ${\cal J_+}, {\cal J_-}$, and ${{\cal J}_3}$ and  
may be written in {\bf pseudo-diagonal form}
\vspace*{-4mm}
        \begin{equation} \label{Hdef}
                {\mathcal H} = \int_{\mathbb{C}} \! d^2 z \, h(z) \, \ket{z} \bra{z} \; .
        \end{equation}
Here the {\bf contravariant symbol} $h$ of $\cal H$ is a (real-valued) 
function on ${\mathbb{C}}$ which 
may be chosen bounded and continuous, 
and  $d^2z := dz_1 dz_2$ is the Lebesgue measure 
on the Euclidean plane ${\mathbb{R}}^2:={\mathbb{R}} \times {\mathbb{R}} \cong {\mathbb{C}}$. 
In particular, the unit operator $\one$ on ${\mathbb{C}}^{2j+1}$
has the constant $1$ as a contravariant symbol. 

Furthermore, for given $z,z' \in \mathbb{C}$, $t>0$, and $\nu > 0$ let $\mathbb{E}(\bcdot)$
denote the probabilistic expectation with 
respect to the {\bf two-dimensional Brownian bridge}
with diffusion constant $\nu$ starting in $z=b(0)$ and arriving at $z'=b(t)$
a time $t$ later.\cite{Sim79,RoWi87}
As a Gaussian stochastic process with continuous paths 
$\left\{s \mapsto b(s)=b_1(s)+ib_2(s)\right\}_{s \in [0,t]}$ 
on ${\mathbb{R}}^2$
it is characterized by its means  
$
                {\mathbb{E}}(b_k(s)) = z_k + (z'_k-z_k) s/t
$
and covariance functions
$
                {\mathbb{E}}(b_k(r) b_l(s)) - 
                {\mathbb{E}}(b_k(r)) {\mathbb{E}}(b_l(s))
                =  2\nu \delta_{kl} \left( \min\{r,s\} - rs/t \right) 
$; $k,l \in \{1,2\}$; $r,s \in [0,t]$.
With the well-defined Wiener type of path integration 
$
\dmut \big(\bcdot\big) :=
                \exp\{- |z-z'|^2/4t \nu\} \,
                {\mathbb{E}}( \bcdot )/4 \pi t \nu
$ 
the {\bf coherent representation} $\bra{z} \exp\{- it {\mathcal H}\} \ket{z'}$ of
the (unitary) {\bf spin time-evolution operator}  $\exp\{-it {\mathcal H}\}$ 
may, for all $z,z' \in {\mathbb{C}}$ and $t>0$, be expressed as    
an {\bf ultra-diffusive limit} in the following sense
\begin{equation} \label{result}
\fbox{$
\begin{array}{rcl}
\displaystyle \bra{z} \ep{- it {\mathcal H}} \ket{z'} & = & \displaystyle
                 \lim_{\nu \to  \infty} \dmut 
                \exp\left\{4(j+1)\nu   \int_0^t 
                        \! \frac{ds}{(1 + |b(s)|^2)^2}\right\} \\[3mm] 
                && \displaystyle \mkern-60mu \times \exp\left\{(j+1) \int_0^t ds 
                \frac{\dot{b}(s) b^*(s) - \dot{b}^*(s) b(s)}{1+|b(s)|^2}
                 -  i \int_0^t \! ds \, h(b(s))\right\}

\end{array}
$}\; \raisebox{-1cm}{.}
\end{equation}
This is the main result of Ref. 3. 
Here the second integral in the exponent  
is a purely imaginary stochastic (line) integral, which can be understood 
in the sense of Fisk and
Stratonovich and to which one is therefore allowed to apply 
the rules of ordinary calculus, although the 
time derivative {$\dot{b}\;$} does not exist.\cite{RoWi87}\\[2mm]
\noindent
Several comments apply:\\[2mm]
\noindent
$\bullet$
Keeping $t>0$ but replacing $h$ by $-h$ or $-ih$ in \eq{result} 
yields analogous 
path-integral expressions for the coherent representation 
of the inverse spin time-evolution
$\exp\{i t {\mathcal H}\}$ or the 
{\bf spin Boltzmann operator} $\exp\{- t {\mathcal H}\}$.\\[2mm]
\noindent
$\bullet$
The flat-space Wiener-regularized path-integral expression \eq{result}
for the spin time-evolution operator is an alternative to a result 
given in  Ref. 2 relying on spherical Brownian motion.  
Contrary to what one might expect, equation \eq{result} cannot be obtained 
from this result merely by stereographically
projecting the paths from the sphere onto the (extended) Euclidean plane. 
In particular, unlike in Ref. 2,
the regularization used in \eq{result} breaks part of 
the $SU(2)$ symmetry when realized on $\mathbb{C}$.
The full symmetry  is restored only in the limit.\\[2mm]
\noindent
$\bullet$
In order to make contact with the
Wiener-regularized path-integral expression associated with 
the dynamics of a canonical degree of freedom,
also proved in Ref. 2, one has to contract 
the $su(2)$ algebra to the 
Heisenberg-Weyl algebra by taking the {\bf high-spin limit} $j \to \infty$.\cite{ACGT72}
More precisely, in the given (polynomial) spin Hamiltonian
$\cal H$ on ${{\mathbb{C}}^{2j+1}}$, one has to replace ${\cal J_+}, {\cal J_-}$, and ${{\cal J}_3}$
by ${\cal J_+}/\sqrt{2j}, {\cal J_-}/\sqrt{2j}$, and ${{\cal J}_3}+j \one$, respectively. 
If ${\cal H}_j = \int_{\mathbb{C}} d^2z \,  h_j(z) \ket{z}\bra{z}$ 
denotes the resulting operator, one then finds the relation 
\begin{equation} \label{contract}
   \lim_{j \to \infty} \frac{\pi}{2j} \bra{z/\sqrt{2j}} \,\ep{- i t {\cal H}_j}\, \ket{ z'/\sqrt{2j}}
     = \langle\!\langle z| \,\ep{- i t {\mathsf H}}\, |z'\rangle\!\rangle
\end{equation}
where $|z\rangle\!\rangle \in \mbox{\rm L}^2(\mathbb R)$ is a normalized 
canonical coherent vector 
and the Hamiltonian $\mathsf H$ on $\mbox{\rm L}^2(\mathbb R)$, the Hilbert space of 
Lebesgue square-integrable complex-valued functions 
on the real line $\mathbb R$, is defined by
$
   {\mathsf H} := \int_{\mathbb{C}} \! \frac{d^2z}{\pi}\, {\mathsf h} (z) |z\rangle\!\rangle \langle\!\langle z| 
$ with $
{\mathsf h}(z) := \lim_{j \to \infty} h_j(z/\sqrt{2j}) 
$.
By using \eq{result} for the pre-limit expression in \eq{contract},
suitably rescaling the Brownian bridge,
and interchanging the order of the limits $j \to \infty$ and 
$\nu \to \infty$, one arrives at the path-integral formula
\vspace*{-0.5mm}
\begin{eqnarray}
\langle\!\langle z| \,\ep{- it {\mathsf H}}\, |z'\rangle\!\rangle  = 
 \pi \lim_{\nu \to  \infty} \ep{2t\nu} \dmut \mkern200mu &&  \nonumber \\
 \times \exp\left\{\half \int_0^t ds 
                \left[\dot{b}(s) b^*(s) - \dot{b}^*(s) b(s)\right]
    - i \int_0^t \! ds \, {\mathsf h}(b(s))\right\} \; ,&&
\end{eqnarray}\\*[-2mm]
in agreement with Eq. (1.3) in  Ref.\ 2, see also Ref. 11.\\[2mm]
\noindent
$\bullet$
With regard to some of the symbolic path-integral expressions for spin systems
frequently encountered
in the literature, see for example  Refs.\ 4-7, it might be
illuminating to recognize certain formal similarities between these expressions and the above result
\eq{result}. While the kinematical and dynamical terms
in the exponents of all the corresponding path integrands 
are essentially the same, only the above result is based on a
genuine path measure, namely  
$\Idmut \exp\{{4(j+1) \nu \int_0^t ds (1 + \abs{b(s)}^2)^{-2}}\}$, but requires
taking the limit $\nu \to \infty$. Here, the Wiener type of measure 
$\Idmut$ is often symbolically written
as $\delta^2 b \, \delta(b(0)-z)
\delta(b(t) - z') \exp\{ - \frac{1}{4 \nu} \int_0^t ds | \dot b(s)|^2\}$
or similarly.\\
%
%

\section{Sketch of the proof}
\vspace*{-2mm}
%
Adapting ideas of Ref. 2, the key to the 
proof of \eq{result} is to identify the spin Hilbert
space ${\mathbb{C}}^{2j+1}$ with the $(2j+1)$-dimensional ground-state
eigenspace $E_0 (\mbox{\rm L}^2({\mathbb{R}}^2))$ of the positive Schr\"odinger operator 
\vspace*{-1mm}
\begin{equation}
R:= \left(i\frac{\partial}{\partial z_1}+ \frac{2(j+1)z_2}{1+|z|^2}\right)^2+
\left(i\frac{\partial}{\partial z_2}- \frac{2(j+1)z_1}{1+|z|^2}\right)^2-
\frac{4(j+1)}{(1+|z|^2)^2}
\end{equation}\\[-2mm]
acting on the Hilbert space
$\mbox{\rm L}^2({\mathbb{R}}^2)$. 
The claimed dimensionality of
$R$'s ground-state projector $E_0$ follows from the 
Aharonov-Casher theorem on zero-energy eigenstates.\cite{CFKS87}
By Taylor expanding about $t=0$ one may now check that the coherent 
representation of the spin 
time-evolution operator may be written as
\begin{equation}
\bra{z}e^{-it{\cal H}}\ket{z'}= E_0 e^{-it E_0 H E_0}(z,z') \, .
\end{equation}
Here the right-hand side denotes the continuous integral kernel 
(in other words position representation) of the operator $E_0 \exp\{-it E_0 H E_0\}$, 
where the bounded multiplication operator $H$ on $\mbox{\rm L}^2({\mathbb{R}}^2)$ is defined by 
$(H \psi)(z) := h(z) \psi(z)$
for all $\psi \in  \mbox{\rm L}^2({\mathbb{R}}^2)$.
It then remains to show the pointwise identity
\begin{equation} \label{limkern}
E_0 e^{-it E_0 H E_0}(z,z') 
= \lim_{\nu \rightarrow \infty} e^{- t(\nu R +i H)}(z,z') \, ,
\end{equation}\\[-0.5mm]
since the pre-limit expressions in \eq{limkern} and  \eq{result} 
coincide by the Feynman-Kac-It\^o formula.\cite{Sim79,BHL98} 
Finally, to prove \eq{limkern} one uses the strong convergence 
$\lim_{\nu \rightarrow \infty} \exp{\{- t \nu R\}} = E_0$ (for all $t>0$) in
the Duhamel-Dyson-Phillips perturbation expansion of 
$\exp{\{-t(\nu R + iH)\}}(z,z')$.
For details see Ref. 3.
\vspace*{-1.8mm}
{\scriptsize


}

\end{document}